\documentclass[prd,superscriptaddress,notitlepage,nofootinbib,twocolumn]{revtex4-2}
\usepackage[T1]{fontenc}
\usepackage[utf8]{inputenc}
\usepackage[english]{babel}
\usepackage{graphicx,microtype,amsfonts,amssymb,amstext,amsmath,mathtools,physics}
\usepackage[hidelinks]{hyperref}

\begin{document}
\title{Pre-geometric Black Holes with Spherical Symmetry}

\author{Giuseppe Meluccio}
    \email{giuseppe.meluccio-ssm@unina.it}
    \affiliation{Scuola Superiore Meridionale, Largo San Marcellino 10, Napoli 80138, Italy}
    \affiliation{INFN Sezione di Napoli, Complesso Universitario di Monte Sant'Angelo, Edificio 6, Via Cintia 21, Napoli 80126, Italy}
    \affiliation{Department of Physics, Loyola Marymount University, 1 LMU Drive, Los Angeles 90045, California, USA}

\author{Egor Pluzhnikov}
    \email{ea.pluzhnikov@outlook.com}
    \affiliation{Faculty of Space Research, Lomonosov Moscow State University, Russia}

\begin{abstract}
    Pre-geometric Gravity investigates the non-metric features of spacetime near the Planck scale, from which the gravitational interaction arises as a low-energy effective theory via a mechanism of spontaneous symmetry breaking. With a view of amending the classical singularities found in black hole geometries, in this work we study pre-geometric black hole configurations with spherical symmetry, and present exact solutions for different pre-geometric theories in vacuum via the correspondence principle. These solutions are the pre-geometric analogue of the (anti-)de Sitter--Schwarzschild metric, from which such type of black hole geometry is thus expected to emerge. In one case, the pre-geometric solution also exhibits an (anti-)de Sitter core of Planckian size. The main result is supplemented with an analysis of the pre-geometric version of Birkhoff's theorem.
\end{abstract}

\maketitle

\section{Introduction}\label{sec:introduction}
The idea of pre-geometry is that gravity can be considered as an emergent phenomenon stemming from some more fundamental physical principles \cite{akama:pregeometry,akama:topological,wetterich:spinors,wetterich:cosmology,wetterich:pregeometry,akama:pregeometric,floreanini:topological}, and in particular the gauge principle \cite{ivanenko:gauge,sardanashvily:gauge,hehl:metric,blagojevic:gravitation,blagojevic:gauge,leclerc:higgs,randono:gauge,ponomarev:gauge,cabral:gauge,celada:bf,de-pietri:plebanski,buffenoir:hamiltonian,durka:hamiltonian} and a gravitational version of the Higgs mechanism \cite{adler:einstein,alexander:higgs,percacci:higgs,volovik:phase}. Pre-geometric theories are formulated as gauge theories for a suitable extension of the Lorentz group $SO(1,3)$, typically the de Sitter (dS) group $SO(1,4)$ or the anti-de Sitter (adS) group $SO(2,3)$ \cite{macdowell:unified,chamseddine:supergravity,stelle:de-sitter,stelle:spontaneously,wilczek:gauge,westman:cartan,westman:dynamical,westman:introduction,pagels:gravitational,mccarthy:surface,mielke:spontaneously,tresguerres:dynamically,wise:cartan,lisi:unification,smolin:holographic,castro:anti,gallagher:pregeometric,palumbo:emergent,chirco:gravity,salgado:gauge}. The symmetries of these gauge groups are defined by the internal metric $\eta=\mathrm{diag}(-,+,+,+,\varepsilon)$, where $\varepsilon=\pm1$ for $SO(1,4)$ or $SO(2,3)$ respectively. Via an appropriate mechanism of spontaneous symmetry breaking (SSB), these gauge groups reduce to $SO(1,3)$ and the internal space metric $\eta$ becomes the Minkowskian one with signature $(-,+,+,+)$. As a consequence of the SSB, the original (a)dS gauge connection of the pre-geometric theory then decomposes into its Lorentzian and tetrad parts, effectively giving rise to the Einstein--Cartan theory, inclusive of the cosmological constant term \cite{addazi:pre-geometry,addazi:hamiltonian,meluccio:pregeometric,addazi:emergence,addazi:topological,addazi:holographic,addazi:solution,addazi:prospective,capozziello:origin,addazi:quantization,addazi:meta-geometric,addazi:bf,addazi:sector,addazi:teleparallel}. It is \emph{only} at this point that a spacetime metric emerges as an effective quantity, otherwise the pre-geometric spacetime is to be viewed as endowed only with topological and differential structures before the SSB, in what is definitely a more primitive concept of spacetime than the one established by Einstein via a metric structure. The energy scales, the fundamental principles and the geometric interpretation of Einstein gravity are thus all emergent in this picture. The pre-geometric SSB is caused by the interaction of the (a)dS gauge potential $A$ with a Higgs-like scalar field $\phi$ in the fundamental representation of the extended gauge group, and happens at an energy scale which is expected to be Planckian from quantum gravity arguments. As discussed in Ref.\ \cite{addazi:pre-geometry}, the main examples of pre-geometric theories are the Wilczek (W) action \cite{wilczek:gauge} and the MacDowell--Mansouri (MM) action \cite{macdowell:unified}, and a possible (though not unique) way to achieve a dynamical realisation of the SSB is through a symmetry-breaking potential for the Higgs-like field (for further discussions, see for example Refs.\ \cite{addazi:pre-geometry,addazi:topological,addazi:holographic,addazi:solution}).

Some predictions of Pre-geometric Gravity for cosmology were analysed, for example, in Refs.\ \cite{meluccio:pregeometric,addazi:topological,addazi:holographic,addazi:solution}; these include the possible resolution of the big bang singularity and of the cosmological constant problem. More broadly, pre-geometric theories envisage a phase transition for the ultra-high-energy behaviour of the gravitational interaction which could effectively remove \emph{any} classical singularity of General Relativity (GR): near the Planck scale, the geometric structure of spacetime would simply be superseded by a non-metric scenario characterised by the dynamics of the pre-geometric fields $A$ and $\phi$, in which the very notion of spacetime singularity would cease to make sense. In crossing the Planck scale, in essence, spacetime itself undergoes a phase transition from a geometric, gravitational background to a pre-geometric, non-metric one. Classical singularities of metric solutions would then come to be seen as mere artefacts of extrapolating GR's formalism beyond its domain of applicability as a low-energy effective theory. For these reasons, in this article we focus on the physical consequences of Pre-geometric Gravity for another case study of GR's singularities, i.e.\ black holes. We start from the most simple case of a spherically symmetric and uncharged black hole, which in GR is described by the (anti-)de Sitter--Schwarzschild solution: following the methodology employed in Ref.\ \cite{meluccio:pregeometric} for the cosmological case, we prove that its pre-geometric analogue is an exact solution of the pre-geometric equations of motion in vacuum. Furthermore, the evolution of the gravitational Higgs field $\phi$ can vary depending on the specific pre-geometric theory under consideration (W or MM), leading to different realisations of the SSB mechanism. In addition to that, we also discuss a proof of the pre-geometric Birkhoff's theorem, thereby showing that Birkhoff's fundamental result for GR can naturally be extended to the pre-geometric phase of a spherically symmetric spacetime.

The notation employed in this paper is the same as that of Ref.\ \cite{meluccio:pregeometric}. Uppercase Latin letters $A,B,C$ etc.\ from $0$ to $4$ and lowercase Latin letters $a,b,c$ etc.\ from $0$ to $3$ represent tangent space (or internal) indices in the unbroken and the spontaneously broken phases respectively. Greek letters $\lambda,\mu,\nu$ etc.\ from $0$ to $3$ indicate spacetime (or external) indices. Moreover, the symbols dot ($\,\dot{}\,$), prime ($'$) and tilde ($\,\tilde{}\,$) denote differentiation with respect to the coordinates $t$, $r$ and $\theta$ respectively. Natural units are assumed throughout the work.

This article is organised as follows. The remainder of Sec.\ \ref{sec:introduction} consists of two subsections, with the first one summarising the main notions about spherically symmetric spacetimes in Einstein gravity and the second one outlining the formalism of pre-geometric theories in vacuum. Secs.\ \ref{sec:spherical} and \ref{sec:birkhoff} then present the main results of this study, respectively about spherically symmetric pre-geometries in vacuum and the pre-geometric Birkhoff's theorem. Finally, Sec.\ \ref{sec:conclusions} provides some concluding remarks as well as directions for future research.

\subsection{Spherically symmetric spacetimes\\in the Palatini formalism}
The geometry of spacetime is described by the metric $g$ and the affine connection $\Gamma$, which evolve according to the field equations of a given gravitational theory. In the Palatini formalism, these two quantities are independent variables -- except for the case of the Levi-Civita connection,
\begin{equation}
    \mathring\Gamma_{\mu\nu}^\lambda=\frac{1}{2}g^{\lambda\rho}(\partial_\mu g_{\rho\nu}+\partial_\nu g_{\mu\rho}-\partial_\rho g_{\mu\nu}),
\end{equation}
which is the unique symmetric (torsionless) and metric-compatible affine connection allowed on a pseudo-Riemannian spacetime.

The Einstein--Cartan theory is a generalisation of GR that employs the Palatini formalism and allows for a torsionful affine connection:
\begin{multline}\label{eq:EC}
    S_\textup{EC}[g,\partial g,\Gamma,\partial\Gamma]\\
    =\frac{1}{16\pi G}\int\sqrt{-g}(g^{\mu\nu}R_{\mu\lambda\nu}^\lambda[\Gamma,\partial\Gamma]-2\Lambda)\,d^4x,
\end{multline}
where $G$ and $\Lambda$ are the gravitational and the cosmological constants respectively, $g$ is the determinant of the metric and $R$ is the curvature tensor. In this theory, the affine connection coincides with the Levi-Civita one whenever fermions are not present, otherwise the torsion of spacetime manifests itself precisely via the non-symmetric character of a more general affine connection. Torsion vanishes in vacuum, therein reducing the Einstein--Cartan theory to GR.

By adopting a description that hinges on tangent space coordinates, the metric and the affine connection can be expressed in an equivalent way in terms of tetrads $e$ and the spin connection $\omega$, via the relations
\begin{gather}
    g_{\mu\nu}=e_\mu^ae_\nu^b\eta_{ab},\\
    \omega_\mu^{ab}=e_\lambda^a\partial_\mu e^{b\lambda}+e_\lambda^a\Gamma_{\mu\nu}^\lambda e^{b\nu},
\end{gather}
with $\eta$ being the Minkowski metric. Using these alternative variables, the Einstein--Cartan theory \eqref{eq:EC} can be recast as
\begin{equation}\label{eq:EC2}
    S_\textup{EC}[e,\partial e,\omega,\partial\omega]=\frac{1}{16\pi G}\int e(e^\mu_ae^\nu_bR_{\mu\nu}^{ab}[\omega,\partial\omega]-2\Lambda)\,d^4x,
\end{equation}
where $e=\sqrt{-g}$ is the tetrad determinant.

In spherical coordinates $(t,r,\theta,\varphi)$, the most general static metric with spherical symmetry can be expressed as
\begin{equation}\label{eq:sss-metric}
    {ds}^2=-f^2(r){dt}^2+g^2(r){dr}^2+r^2{d\theta}^2+r^2\sin^2\theta{d\varphi}^2,
\end{equation}
with $f$ and $g$ being independent functions of the radial coordinate. The nonzero components of the corresponding tetrads and Levi-Civita spin connection are respectively
\begin{subequations}\label{eq:static-spherical-symmetry}
    \begin{gather}
        e_0^0=f(r),\qquad e_1^1=g(r),\qquad e_2^2=r,\qquad e_3^3=r\sin\theta,\\
        \mathring\omega_0^{01}=\frac{f'(r)}{g(r)},\qquad\mathring\omega_2^{12}=-\frac{1}{g(r)},\nonumber\\
        \mathring\omega_3^{13}=-\frac{\sin\theta}{g(r)},\qquad\mathring\omega_3^{23}=-\cos\theta.
    \end{gather}
\end{subequations}
If the static condition is lifted, then the most general metric with spherical symmetry is
\begin{equation}\label{eq:tss-metric}
    {ds}^2=-f^2(t,r){dt}^2+g^2(t,r){dr}^2+r^2{d\theta}^2+r^2\sin^2\theta{d\varphi}^2,
\end{equation}
with $f$ and $g$ being independent functions of the time and radial coordinates. The nonzero components of the corresponding tetrads and Levi-Civita spin connection\footnote{For a broader discussion on spherically symmetric geometries in the Palatini formalism, the interested reader is referred to Refs.\ \cite{minkevich:spherically-symmetric,hohmann:metric-affine}.} are then respectively
\begin{subequations}\label{eq:non-static-spherical-symmetry}
    \begin{gather}
        e_0^0=f(t,r),\qquad e_1^1=g(t,r),\qquad e_2^2=r,\qquad e_3^3=r\sin\theta,\\
        \mathring\omega_0^{01}=\frac{f'(t,r)}{g(t,r)},\qquad\mathring\omega_1^{01}=\frac{\dot g(t,r)}{f(t,r)},\qquad\mathring\omega_2^{12}=-\frac{1}{g(t,r)},\nonumber\\
        \mathring\omega_3^{13}=-\frac{\sin\theta}{g(t,r)},\qquad\mathring\omega_3^{23}=-\cos\theta.
    \end{gather}
\end{subequations}

\begin{figure}
	\centering
	\includegraphics[scale=0.5]{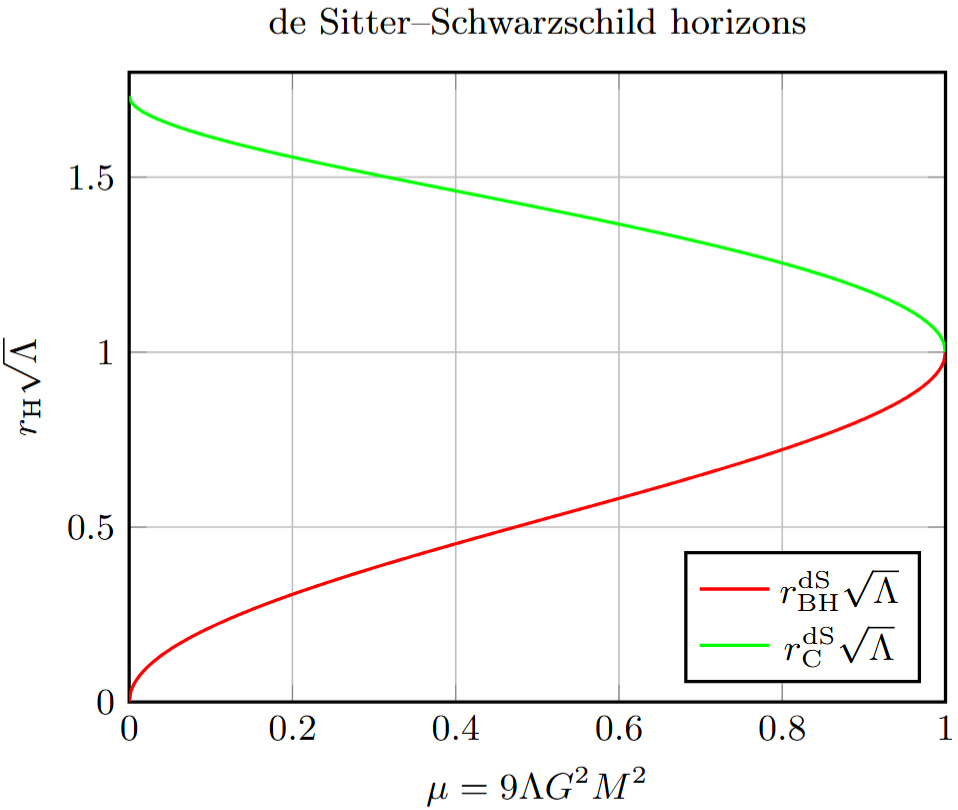}
	\caption{Dimensionless horizons $r_\textup{H}\sqrt{\Lambda}$ of the de Sitter--Schwarzschild spacetime ($\Lambda>0$) as a function of the dimensionless mass parameter $\mu$. Note that $r_\textup{BH}^\textup{dS}=r_\textup{C}^\textup{dS}$ in the Nariai limit $\mu=1$.\label{fig:dS}}
\end{figure}

\begin{figure}
	\centering
	\includegraphics[scale=0.5]{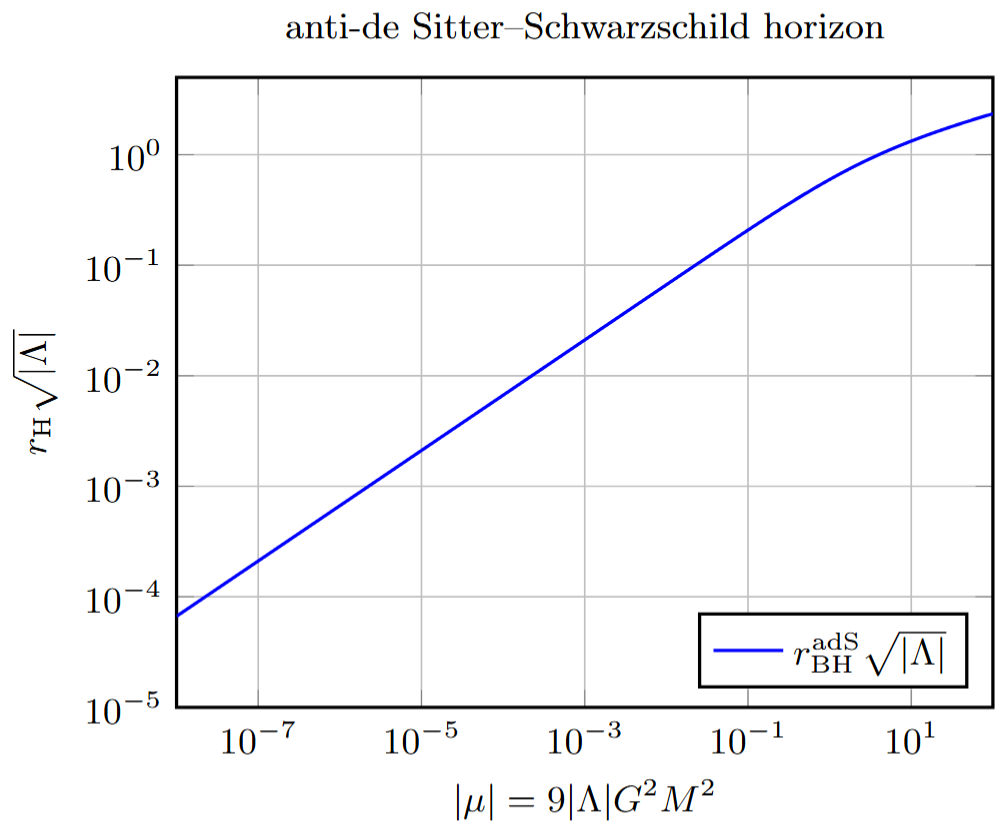}
	\caption{Dimensionless horizon $r_\textup{H}\sqrt{\abs{\Lambda}}$ of the anti-de Sitter--Schwarzschild spacetime ($\Lambda<0$) as a function of the dimensionless mass parameter $\abs{\mu}$.\label{fig:adS}}
\end{figure}

The (anti-)de Sitter--Schwarzschild spacetime is the unique static metric with spherical symmetry that is solution of the vacuum Einstein field equations with a (negative/)positive cosmological constant. It is a special case of Eq.\ \eqref{eq:sss-metric} with
\begin{equation}\label{eq:adS-S}
    f^2(r)=\frac{1}{g^2(r)}=1-\frac{2GM}{r}-\frac{\Lambda}{3}r^2,
\end{equation}
where $M$ is the mass of the central matter source and $\Lambda=\varepsilon\abs{\Lambda}$. The horizon structure of this metric depends on the value of $\varepsilon$. For $\varepsilon=+1$, the global topology of the de Sitter--Schwarzschild spacetime hinges on the value of the dimensionless parameter $\mu\equiv9\varepsilon\abs{\Lambda}G^2M^2$:
\begin{itemize}
    \item for $\mu<1$ ($M<\frac{1}{3G\sqrt{\Lambda}}$), two distinct horizons exist and are given by the formula
    \begin{equation}
        \begin{split}
            r_\textup{H}^{(k)}&=\frac{2}{\sqrt{\Lambda}}\cos\biggl[\frac{1}{3}\arccos(-\sqrt\mu)-\frac{2}{3}k\pi\biggr]\\
            &=\frac{2}{\sqrt{\Lambda}}\cos\biggl(\frac{\pi}{3}-\frac{1}{3}\arccos\sqrt\mu-\frac{2}{3}k\pi\biggr),
        \end{split}
    \end{equation}
    where $r_\textup{H}^{(k=1)}\equiv r_\textup{BH}^\textup{dS}$ is the black hole event horizon and $r_\textup{H}^{(k=0)}\equiv r_\textup{C}^\textup{dS}$ the cosmological horizon (with $r_\textup{BH}^\textup{dS}<r_\textup{C}^\textup{dS}$);
    \item for $\mu=1$ ($M=\frac{1}{3G\sqrt{\Lambda}}$ which, upon restoring a factor $c^2$ at the numerator, amounts to $M\approx2\cdot10^{22}\,M_\odot$), the two horizons coincide (Nariai limit);
    \item for $\mu>1$ ($M>\frac{1}{3G\sqrt{\Lambda}}$), no horizons exist and the singularity at $r=0$ is naked.
\end{itemize}
Astrophysical black holes in our universe fall into the category with $\mu\ll1$. On the other hand, for $\varepsilon=-1$, the global topology of the anti-de Sitter--Schwarzschild spacetime is independent of the value of $\mu$ and always possesses one and only one horizon; this is a black hole event horizon located at
\begin{equation}
    r_\textup{H}=\frac{-1+(\sqrt{-\mu}+\sqrt{1-\mu})^{2/3}}{\sqrt{-\Lambda}(\sqrt{-\mu}+\sqrt{1-\mu})^{1/3}}\equiv r_\textup{BH}^\textup{adS},
\end{equation}
while no cosmological horizon exists. Two plots for the different event horizons of the (anti-)de Sitter--Schwarzschild metric are presented in Figs.\ \ref{fig:dS} and \ref{fig:adS}.


We conclude this subsection with a recollection of a key result in Einstein gravity: Birkhoff's theorem states that any spherically symmetric solution of GR's vacuum field equations is static; such solution is unique and is the (anti-)de Sitter--Schwarzschild spacetime, for which the Schwarzschild spacetime represents the special case ($\Lambda=0$) that is also asymptotically flat.

\subsection{The pre-geometric vacuum}\label{subsec:pre-geometric-vacuum}
The pre-geometric framework is formulated on a four-dimensional differentiable manifold $M$ as a gauge theory for a gauge group $G$, usually taken to be either the de Sitter (ds) group $SO(1,4)$ or the anti-de Sitter (adS) group $SO(2,3)$. The fundamental gauge group $G$ admits a decomposition into a subgroup $H$ and a complementary sector $P$, such that $H\subseteq G=H \oplus P$; in the cases of physical interest, a mechanism of SSB via a Higgs-like scalar field $\phi$ yields $H=SO(1,3)$. The gauge potential of the theory is denoted as $A\in\Omega^1(M,\mathfrak{g})$, where $\mathfrak{g}=\mathrm{Lie}(G)$. Note that $\mathfrak{g}=\mathfrak{h}\oplus \mathfrak{p}$ due to the decomposition stated above. The associated gauge curvature $F=dA+A\wedge A\in\Omega^2(M,\mathfrak{g})$ is given in components by
\begin{equation}
    F_{\mu\nu}^{AB}=2\partial_{[\mu}A_{\nu]}^{AB}+2A_{C[\mu}^AA_{\nu]}^{CB},
\end{equation}
where $A^{A}_{B\mu}\equiv A^{A}_{\phantom{B}B\mu}=\eta_{BC} A^{AC}_\mu$. The corresponding gauge covariant derivative is denoted as $\nabla$ and its action on $\phi$ is given in components by
\begin{equation}
    \nabla_\mu\phi^A=\partial_\mu\phi^A+A_{B\mu}^A\phi^B.
\end{equation}

The SSB $G\rightarrow H$ is dynamically realised \cite{addazi:pre-geometry} when the Higgs-like field acquires a nonzero vacuum expectation value (VEV) as $\langle\phi^A\rangle=v\delta^A_4$. This leads to the decomposition of the gauge potential as follows:
\begin{equation}
    A_\mu^{ab}\xrightarrow{SSB}\omega_\mu^{ab},\qquad A_\mu^{a4}\xrightarrow{SSB}me_\mu^a,
\end{equation}
where $m$ is a mass parameter (this allows to extract dimensionless tetrad fields from the dimensionful gauge potential).

In this paper, we will consider two pre-geometric theories of gravity, namely the Wilczek (W) one \cite{wilczek:gauge} and the MacDowell--Mansouri (MM) one \cite{macdowell:unified}. The action of the W theory is obtained by integrating the Lagrangian density
\begin{equation}\label{eq:L-W}
    \mathcal{L}_\textup{W}=k_\textup{W}\epsilon_{ABCDE}\epsilon^{\mu\nu\rho\sigma}F_{\mu\nu}^{AB}\nabla_\rho\phi^C\nabla_\sigma\phi^D\phi^E
\end{equation}
over the whole spacetime manifold, with the corresponding equations of motion in vacuum\footnote{As noted in Ref.\ \cite{meluccio:pregeometric}, it is possible that, even in the absence of matter, additional terms should be taken into account for the vacuum dynamics of pre-geometric spacetimes, such as kinetic or potential terms for $A$ and $\phi$. These terms are neglected in the present analysis for simplicity, but also because their specific form hinges on the specific quantisation scheme to be implemented (see Refs.\ \cite{wilczek:gauge,addazi:pre-geometry,addazi:hamiltonian,meluccio:pregeometric,addazi:topological,addazi:emergence,addazi:holographic,addazi:solution} for further discussions).} being
\begin{subequations}\label{eq:EoM-W}
    \begin{gather}
        \epsilon^{\mu\nu\rho\sigma}[\epsilon_{ABCD[E}\eta_{F]G}(F_{\mu\nu}^{AB}\phi^G-2A_\mu^{AG}\nabla_\nu\phi^B)\nabla_\rho\phi^C\phi^D\nonumber\\
        +\epsilon_{ABCEF}\partial_\rho(\nabla_\mu\phi^A\nabla_\nu\phi^B\phi^C)]=0,\label{eq:EoM-W-A}\\
        \epsilon^{\mu\nu\rho\sigma}[(2\epsilon_{ABCDF}A_{E\sigma}^D\phi^F+\epsilon_{ABCDE}\nabla_\sigma\phi^D)F_{\mu\nu}^{AB}\nabla_\rho\phi^C\nonumber\\
        +2\epsilon_{ABCDE}\partial_\sigma(F_{\mu\nu}^{AB}\nabla_\rho\phi^C\phi^D)]=0.\label{eq:EoM-W-phi}
    \end{gather}
\end{subequations}
Similarly, the Lagrangian density for the MM theory is
\begin{equation}\label{eq:L-MM}
    \mathcal{L}_\textup{MM}=k_\textup{MM}\epsilon_{ABCDE}\epsilon^{\mu\nu\rho\sigma}F_{\mu\nu}^{AB}F_{\rho\sigma}^{CD}\phi^E
\end{equation}
and its equations of motion in vacuum are
\begin{subequations}\label{eq:EoM-MM}
    \begin{gather}
        \epsilon^{\mu\nu\rho\sigma}[2\epsilon_{ABCD[E}A_{F]\mu}^AF_{\nu\rho}^{BC}\phi^D-\epsilon_{ABCEF}\partial_\rho(F_{\mu\nu}^{AB}\phi^C)]=0,\label{eq:EoM-MM-A}\\
        \epsilon_{ABCDE}\epsilon^{\mu\nu\rho\sigma}F_{\mu\nu}^{AB}F_{\rho\sigma}^{CD}=0.\label{eq:EoM-MM-phi}
    \end{gather}
\end{subequations}
Note that both pre-geometric theories rely on three undetermined parameters: the energy scale $m$, the VEV $v$ (which is to be determined dynamically via the symmetry-breaking potential) and the coupling constant $k_\textup{W}$ or $k_\textup{MM}$. It is some combinations of these parameters that give rise, in the spontaneously broken phase, to the (reduced) Planck mass and the cosmological constant. Indeed, both pre-geometric theories recover the Einstein--Cartan theory \eqref{eq:EC2} as the effective dynamics of the emergent gravitational degrees of freedom (with the MM theory recovering a Gauss--Bonnet term as well \cite{addazi:pre-geometry}). For the W theory this requires the identifications
\begin{equation}\label{eq:emergent-constants-W}
	M_\textup{P}^2\equiv-8k_\textup{W}v^3m^2,\qquad\Lambda\equiv6\varepsilon m^2,
\end{equation}
while for the MM theory the correct identities are
\begin{equation}\label{eq:emergent-constants-MM}
	M_\textup{P}^2\equiv32\varepsilon k_\textup{MM}vm^2,\qquad\Lambda\equiv3\varepsilon m^2.
\end{equation}

\section{Spherically symmetric pre-geometries in vacuum}\label{sec:spherical}
In this section we search for exact solutions of the pre-geometric equations of motion \eqref{eq:EoM-W} and \eqref{eq:EoM-MM} that are spherically symmetric and static. To do so, we resort to the correspondence principle \cite{meluccio:pregeometric}, which maintains that the SSB of any pre-geometric solution must match the corresponding geometric solution of the Einstein--Cartan theory satisfying the same symmetry requirements. The viewpoint here is to treat the unbroken phase as a possible UV completion of the spontaneously broken one, as the phase transition from a pre-geometric to a geometric universe is expected to happen near the Planck scale \cite{addazi:pre-geometry}.\footnote{Given the expected Planckian mass of the excitations $\rho$ of the Higgs-like field $\phi$, in this work such fluctuations will be considered as effectively frozen out in the IR limit where the Einstein--Cartan theory emerges. This justifies writing $\phi^A\xrightarrow{SSB}v\delta^A_4$ instead of $\phi^A\xrightarrow{SSB}(v+\rho)\delta^A_4$, with the latter expression representing the value of the Higgs-like field in the unitary gauge when its fluctuations $\rho$ are excited near the Planck scale (see also Ref.\ \cite{meluccio:pregeometric}).} Starting from the geometry \eqref{eq:static-spherical-symmetry} of the most general static metric with spherical symmetry, we can thus reverse-engineer an ansatz for the corresponding pre-geometric spacetime:
\begin{equation}\label{eq:ansatz1}
    \begin{split}
        A_0^{04}&\equiv\alpha(r),\qquad A_1^{14}\equiv\beta(r),\qquad A_2^{24}\equiv\gamma(r),\\
        A_3^{34}&\equiv\delta(r,\theta),\qquad A_0^{01}\equiv\tau(r),\qquad A_2^{12}\equiv\xi(r),\\
        A_3^{13}&\equiv\kappa(r,\theta),\qquad A_3^{23}\equiv\zeta(\theta),
    \end{split}
\end{equation}
with all other components of the gauge potential (which are not obtained from its antisymmetry $A_\mu^{AB}=-A_\mu^{BA}$) being zero. The eight functions of $r$ or $\theta$ introduced in the ansatz \eqref{eq:ansatz1} must be complemented with a ninth one for the Higgs-like field, which will be expressed in the unitary gauge (as done in Refs.\ \cite{addazi:pre-geometry,meluccio:pregeometric}): $\phi^A\equiv\Phi(r)\delta_4^A$. Therefore, at least nine independent equations are needed in order to find exact pre-geometric solutions subject to the symmetries under consideration.

\subsection{Static and spherical symmetry in the Wilczek theory}
Plugging the ansatz \eqref{eq:ansatz1} into Eqs.\ \eqref{eq:EoM-W-A}, the equations of motion for the gauge potential $A$ in the W theory are found to be
\begin{subequations}\label{eq:EoMforAbyW1}
    \begin{gather}
        \beta\Phi(\gamma\kappa+\delta\xi)+3\gamma\delta\Phi'+\Phi(\gamma\delta)'=0,\\
        \gamma\zeta+\tilde\delta=0,\\
        \beta(6\varepsilon\gamma\delta+\kappa\xi-\tilde\zeta)-\delta\xi'-\gamma\kappa'=0,\\
        \tau(\gamma\kappa+\delta\xi)-\alpha(6\varepsilon\gamma\delta+\kappa\xi-\tilde\zeta)=0,\\
        \xi\zeta+\tilde\kappa=0,\\
        \delta\Phi(\beta\tau-\alpha')-\alpha(\beta\kappa\Phi+3\delta\Phi'+\Phi\delta')=0,\\
        \beta\kappa\tau-\delta\tau'-\alpha(6\varepsilon\beta\delta-\kappa')=0,\\
        \alpha(\beta\xi\Phi+\Phi\gamma'+3\gamma\Phi')-\gamma\Phi(\beta\tau-\alpha')=0,\\
        \beta\xi\tau-\gamma\tau'-\alpha(6\varepsilon\beta\gamma-\xi')=0.
    \end{gather}
\end{subequations}
Similarly, the equations of motion for the Higgs-like field $\phi$ descending from the Eqs.\ \eqref{eq:EoM-W-phi} of the W theory are
\begin{subequations}\label{eq:EoMforPhibyW1}
    \begin{gather}
        \Phi(4\varepsilon\gamma\delta+\kappa\xi-\tilde\zeta)\alpha'-\tau\Phi[\kappa\gamma'+\xi\delta'+\beta(4\varepsilon\gamma\delta+3\kappa\xi-\tilde\zeta)]\nonumber\\
        -3\tau(\gamma\kappa+\delta\xi)\Phi'+\alpha\{4\varepsilon\Phi[\beta(\gamma\kappa+\delta\xi)+(\gamma\delta)']\nonumber\\
        +3(\kappa\xi-\tilde\zeta+3\varepsilon\gamma\delta)\Phi'\}=0,\\
        [\Phi(\alpha'-\beta\tau)+3\alpha\Phi'](\xi\zeta+\tilde\kappa)\nonumber\\
        +\Phi(\gamma\zeta+\tilde\delta)(\tau'+4\varepsilon\alpha\beta)=0,\\
        \beta\tau(\gamma\kappa+\delta\xi)-\gamma\delta\tau'\nonumber\\
        -\alpha[\beta(6\varepsilon\gamma\delta+\kappa\xi-\tilde\zeta)-\delta\xi'-\gamma\kappa']=0.
    \end{gather}
\end{subequations}

A class of solutions to the above set of coupled differential equations is given, in terms of the function $\alpha$, by
\begin{equation}\label{eq:class-solutions1}
    \begin{split}
        \beta(r)&=\frac{m^2}{\alpha(r)},\qquad\gamma(r)=mr,\qquad\delta(r,\theta)=\gamma(r)\sin\theta,\\
        \tau(r)&=\frac{\alpha(r)\alpha'(r)}{m^2},\qquad\xi(r)=-\frac{\alpha(r)}{m},\\
        \kappa(r,\theta)&=-\frac{\alpha(r)}{m}\sin\theta,\qquad\zeta(\theta)=-\cos\theta,
    \end{split}
\end{equation}
where different factors of $m$ were included for dimensional reasons (see the discussion in Subsec.\ \ref{subsec:pre-geometric-vacuum}). For the class of solutions \eqref{eq:class-solutions1}, the equations of motion \eqref{eq:EoMforAbyW1} and \eqref{eq:EoMforPhibyW1} reduce respectively to
\begin{subequations}\label{eq:EoMforAbyW2}
    \begin{gather}
        \Phi'=0,\\
        -m^2+6\varepsilon m^4r^2+\alpha^2+2r\alpha\alpha'=0,\\
        r[6\varepsilon m^4+(\alpha')^2]+\alpha(2\alpha'+r\alpha'')=0
    \end{gather}
\end{subequations}
and
\begin{subequations}\label{eq:EoMforPhibyW2}
    \begin{gather}
        (-m^2+3\varepsilon m^4r^2+\alpha^2+2r\alpha\alpha')\Phi'=0,\\
        -m^2+r^2[6\varepsilon m^4+(\alpha')^2]+\alpha^2+r\alpha(4\alpha'+r\alpha'')=0.
    \end{gather}
\end{subequations}
The unique solution of Eqs.\ \eqref{eq:EoMforAbyW2} is
\begin{equation}\label{eq:SolforAbyW}
    \alpha(r)=m\sqrt{1-\frac{\mathcal{C}_1}{r}-2\varepsilon m^2r^2},\qquad\Phi(r)=\text{const.},
\end{equation}
which renders Eqs.\ \eqref{eq:EoMforPhibyW2} irrelevant as $\Phi$ is constant on shell, meaning that $\phi^4$ can no longer be considered as a local degree of freedom under these symmetry constraints. The SSB of the result \eqref{eq:SolforAbyW} yields
\begin{equation}
    f(r)=\sqrt{1-\frac{\mathcal{C}_1}{r}-\frac{\Lambda}{3}r^2}.
\end{equation}
The pre-geometric solution represented by Eqs.\ \eqref{eq:class-solutions1} and \eqref{eq:SolforAbyW} thus reduces exactly to the (anti-)de Sitter--Schwarzschild spacetime (Eqs.\ \eqref{eq:static-spherical-symmetry} and \eqref{eq:adS-S}) in the spontaneously broken phase, provided that the integration constant is identified as $\mathcal{C}_1=2GM$. Observe that, just as in the cosmological case \cite{meluccio:pregeometric}, the W theory exhibits a stabilisation of the pre-geometric scalar field $\phi$ under the assumed symmetries: as the constant value of $\phi$ is identified with its VEV $v$, the SSB of a static and spherically symmetric pre-geometric spacetime is inevitable in the W theory, and necessarily leads to the emergence of the (anti-)de Sitter--Schwarzschild metric.

\subsection{Static and spherical symmetry in the MacDowell--Mansouri theory}
In this subsection we follow the same steps as in the previous one, this time dealing with the MM theory. By applying the ansatz \eqref{eq:ansatz1} to the equations \eqref{eq:EoM-MM-A} and \eqref{eq:EoM-MM-phi} of the MM theory, one obtains the following equations of motion for $A$ and $\phi$ respectively:
\begin{subequations}\label{eq:EoMforAbyMM1}
    \begin{gather}
        \varepsilon\beta\Phi(\gamma\kappa+\delta\xi)+(\kappa\xi-\tilde\zeta)\Phi'+\varepsilon(\gamma\delta\Phi)'=0,\\
        \varepsilon\beta\Phi(\gamma\zeta+\tilde\delta)+(\xi\zeta+\tilde\kappa)\Phi'=0,\\
        \gamma\zeta+\tilde\delta=0,\\
        \delta\xi'+\gamma\kappa'-\beta(3\varepsilon\gamma\delta+\kappa\xi-\tilde\zeta)=0,\\
        \tau(\gamma\kappa+\delta\xi)-\alpha(3\varepsilon\gamma\delta+\kappa\xi-\tilde\zeta)=0,\\
        \xi\zeta+\tilde\kappa=0,\\
        \beta\Phi(\delta\tau-\alpha\kappa)-(\alpha\delta\Phi)'+\varepsilon\kappa\tau\Phi'=0,\\
        \beta\kappa\tau-\delta\tau'-\alpha(3\varepsilon\beta\delta-\kappa')=0,\\
        \beta\Phi(\gamma\tau-\alpha\xi)-(\alpha\gamma\Phi)'+\varepsilon\tau\xi\Phi'=0,\\
        \beta\tau\xi-\gamma\tau'-\alpha(3\varepsilon\beta\gamma-\xi')=0
    \end{gather}
\end{subequations}
and
\begin{subequations}\label{eq:EoMforPhibyMM1}
    \begin{gather}
        (\varepsilon\gamma\delta+\kappa\xi-\tilde\zeta)\alpha'-\kappa\tau\gamma'-\beta\tau(\varepsilon\gamma\delta+3\kappa\xi\tilde\zeta)\nonumber\\
        -\tau\xi\delta'-\varepsilon\alpha[\beta(\gamma\kappa+\delta\xi)+(\gamma\delta)']=0,\\
        (\xi\zeta+\tilde\kappa)(\alpha'-\beta\tau)+(\gamma\zeta+\tilde\delta)(\tau'+\varepsilon\alpha\beta)=0,\\
        \varepsilon\beta\tau(\gamma\kappa+\delta\xi)-\varepsilon\gamma\delta\tau'+\tau'\tilde\zeta-(\kappa\tau\xi)'\nonumber\\
        -\varepsilon\alpha[\beta(3\varepsilon\gamma\delta+\kappa\xi-\tilde\zeta)-\delta\xi'-\gamma\kappa']=0.
    \end{gather}
\end{subequations}

Also in this case, a class of solutions, in terms of the function $\alpha$, is given by Eq.\ \eqref{eq:class-solutions1}, which reduces Eqs.\ \eqref{eq:EoMforAbyMM1} and \eqref{eq:EoMforPhibyMM1} respectively to
\begin{subequations}\label{eq:EoMforAbyMM2}
    \begin{gather}
        (-m^2+\varepsilon m^4r^2+\alpha^2)\Phi'=0,\\
        -m^2+3\varepsilon m^4r^2+\alpha^2+2r\alpha\alpha'=0,\\
        (\varepsilon m^4r+\alpha\alpha')\Phi'=0,\\
        r[3\varepsilon m^4+(\alpha')^2]+\alpha(2\alpha'+r\alpha'')=0
    \end{gather}
\end{subequations}
and
\begin{multline}\label{eq:EoMforPhibyMM2}
    \varepsilon m^6(3\varepsilon m^2r^2-1)+m^2(\varepsilon m^2r^2-1)(\alpha')^2+\alpha^3\alpha''\\
    +\alpha^2[\varepsilon m^4+3(\alpha')^2]+\alpha[4\varepsilon m^4r\alpha'+m^2(\varepsilon m^2r^2-1)\alpha'']=0.
\end{multline}
This time, Eqs.\ \eqref{eq:EoMforAbyMM2} admit two distinct solutions:
\begin{subequations}\label{eq:SolforAbyMM}
    \begin{gather}
        \alpha(r)=m\sqrt{1-\frac{\mathcal{C}_2}{r}-\varepsilon m^2r^2},\qquad\Phi(r)=\text{const.},\label{eq:SolforAbyMMa}\\
        \alpha(r)=m\sqrt{1-\varepsilon m^2r^2},\label{eq:SolforAbyMMb}
    \end{gather}
\end{subequations}
with the function $\Phi$ being constant in the first branch \eqref{eq:SolforAbyMMa} and undetermined in the second one \eqref{eq:SolforAbyMMb}. The SSB of the results \eqref{eq:SolforAbyMM} gives respectively
\begin{subequations}
    \begin{gather}
        f(r)=\sqrt{1-\frac{\mathcal{C}_2}{r}-\frac{\Lambda}{3}r^2},\\
        f(r)=\sqrt{1-\frac{\Lambda}{3}r^2}.
    \end{gather}
\end{subequations}
The solution in the first branch is thus identical to the one found for the W theory in the spontaneously broken phase, giving rise to an emergent (anti-)de Sitter--Schwarzschild metric with a constant Higgs-like field (again, provided that the integration constant is identified as $\mathcal{C}_2=2GM$). As for the solution in the second branch, after the SSB it corresponds to the (anti-)de Sitter spacetime, that is an empty (anti-)de Sitter--Schwarzschild spacetime ($M=0$). This solution is regular at the centre in the sense that no singularities arise in the limit $r\rightarrow0$. For the second branch, it is important to solve the equation of motion for $\phi$ as well, i.e.\ Eq.\ \eqref{eq:EoMforPhibyMM2}; this admits two solutions too, namely
\begin{equation}\label{eq:SolforPhibyMM}
    \alpha(r)=m\sqrt{1-\varepsilon m^2r^2\pm\sqrt{\mathcal{C}_3+\mathcal{C}_4r}},
\end{equation}
which are both compatible with the result \eqref{eq:SolforAbyMMb} provided that the two integration constants are $\mathcal{C}_3=\mathcal{C}_4=0$. Therefore, the solution in the second branch gives rise to an emergent (anti-)de Sitter metric with an undetermined Higgs-like field, whose evolution can thus effectively realise the mechanism of SSB between the pre-geometric and the geometric phases of spacetime in a dynamical way.\footnote{The situation with an undetermined Higgs-like field in the unbroken phase of the MM theory with static and spherical symmetries is the same as the one found for the cosmological pre-geometric solution of the same theory in Ref.\ \cite{meluccio:pregeometric}. From a mathematical point of view, this result is due to the peculiar algebraic nature of $\phi$ in the Lagrangian density \eqref{eq:L-MM}, which reflects on the complete absence thereof in Eq.\ \eqref{eq:EoMforPhibyMM2} and its appearance with a derivative in Eqs.\ \eqref{eq:EoMforAbyMM2} but only as an overall multiplicative factor.}

In conclusion, the static and spherically symmetric spacetime that is predicted by the MM theory is more complex than that of the W theory: it can be thought of as an emergent (anti-)de Sitter--Schwarzschild spacetime (first-branch solution) with a regular pre-geometric (anti-)de Sitter core (second-branch solution) of Planckian size. Rather then being a pathology here, the free evolution of the pre-geometric Higgs field is the feature which allows the unbroken phase to be actually realisable (as opposed to the case of the W theory), at least in the absence of additional terms in the action; therefore, the SSB needs not automatically take place in this scenario, rather it is only the specifics of a possible symmetry-breaking potential that determine its eventual occurrence near the Planck scale. The interpretation of this solution as having a regular pre-geometric core thus stems precisely from the evolution of $\phi$: in the first branch (which happens not to be regular for $r\rightarrow0$), $\phi$ is frozen to a constant value and so the pre-geometric SSB is inevitable, which is the case for low energies / large distances (outside the core); in the second branch (which happens to be regular for $r\rightarrow0$), instead, $\phi$ can fluctuate around its VEV according to pre-geometric dynamics (though still subject to the effects of possible additional terms such as the symmetry-breaking potential), which is the case for high (Planckian) energies / short distances (inside the core).

\section{Pre-geometric Birkhoff's theorem}\label{sec:birkhoff}
We now discuss the pre-geometric Birkhoff's theorem, i.e.\ the statement that any spherically symmetric solution of the pre-geometric field equations in vacuum is static. As done in Sec.\ \ref{sec:spherical}, our starting point is again the correspondence principle. This time, instead of basing our ansatz on the spherically symmetric and static geometry \eqref{eq:static-spherical-symmetry}, we have to look at its non-static version \eqref{eq:non-static-spherical-symmetry}, which translates into the following radial ansatz for a time-dependent pre-geometry of spacetime:
\begin{equation}\label{eq:ansatz2}
    \begin{split}
        A_0^{04}&\equiv\alpha(t,r),\qquad A_1^{14}\equiv\beta(t,r),\qquad A_2^{24}\equiv\gamma(t,r),\\
        A_3^{34}&\equiv\delta(t,r,\theta),\qquad A_0^{01}\equiv\tau(t,r),\qquad A_1^{01}\equiv\sigma(t,r),\\
        A_2^{12}&\equiv\xi(t,r),\qquad A_3^{13}\equiv\kappa(t,r,\theta),\qquad A_3^{23}\equiv\zeta(t,\theta),
    \end{split}
\end{equation}
with $\phi^A\equiv\Phi(t,r)\delta_4^A$. Note that in this case there are ten independent functions of $t$, $r$ and $\theta$ to be determined.

\subsection{Spherical symmetry in the Wilczek theory}
After plugging the ansatz \eqref{eq:ansatz2} into Eqs.\ \eqref{eq:EoM-W}, this time we will not show the equivalent of Eqs.\ \eqref{eq:EoMforAbyW1}--\eqref{eq:EoMforPhibyW1} because they are too complicated; rather, we will focus directly on the class of solutions corresponding to the equivalent of Eq.\ \eqref{eq:class-solutions1}, that is:
\begin{equation}\label{eq:class-solutions2}
    \begin{split}
        \beta(t,r)&=\frac{m^2}{\alpha(t,r)},\qquad\gamma(t,r)=mr,\\
        \delta(t,r,\theta)&=\gamma(t,r)\sin\theta,\qquad\tau(t,r)=\frac{\alpha(t,r)\alpha'(t,r)}{m^2},\\
        \sigma(t,r)&=\frac{\dot\beta(t,r)}{\alpha(t,r)},\qquad\xi(t,r)=-\frac{\alpha(t,r)}{m},\\
        \kappa(t,r,\theta)&=-\frac{\alpha(t,r)}{m}\sin\theta,\qquad\zeta(t,\theta)=-\cos\theta.
    \end{split}
\end{equation}
An important observation to be made here is that the class of solutions \eqref{eq:class-solutions2} reduces to that given by Eq.\ \eqref{eq:class-solutions1} for the static case if $\dot\alpha=0$.

For our purposes, it then suffices to analyse the following relevant equations of motion for the gauge potential $A$ for the class of solutions \eqref{eq:class-solutions2} in the W theory:
\begin{subequations}\label{eq:EoMforAbyW3}
    \begin{gather}
        \Phi'=0,\\
        \dot\Phi=0,\\
        \dot\alpha=0.\label{eq:EoMforAbyW3-Birkhoff}
    \end{gather}
\end{subequations}
In particular, Eq.\ \eqref{eq:EoMforAbyW3-Birkhoff} -- which corresponds to the equations of motion for the gauge potential $A$ with $(\mu,A,B)=(1,0,4)$ -- proves the pre-geometric Birkhoff's theorem in the Wilczek theory for the class of solutions \eqref{eq:class-solutions2}. From this result and the observation made right after Eq.\ \eqref{eq:class-solutions2}, it follows that the solution to the full set of time-dependent equations of motion for $A$ in the W theory (which includes but is not limited to Eqs.\ \eqref{eq:EoMforAbyW3}) coincides with the solution \eqref{eq:SolforAbyW} presented for the static case.

\subsection{Spherical symmetry in the MacDowell--Mansouri theory}
Again, after inserting the ansatz \eqref{eq:ansatz2} into Eqs.\ \eqref{eq:EoM-MM}, we will not show the equivalent of Eqs.\ \eqref{eq:EoMforAbyMM1}--\eqref{eq:EoMforPhibyMM1} because they are too complicated; rather, we will proceed directly to study the class of solutions given by Eq.\ \eqref{eq:class-solutions2}.

The only relevant equation of motion for the Higgs-like field $\phi$ for the class of solutions \eqref{eq:class-solutions2} in the MM theory is
\begin{multline}\label{eq:EoMforPhibyMM3}
    m^2\alpha^4\{\varepsilon m^2[m^2(3\varepsilon m^2r^2-1)+4r\alpha\alpha']\\
    +(\varepsilon m^2r^2-1)[\alpha\alpha''+(\alpha')^2]\}+\alpha^7\alpha''+\alpha^6[\varepsilon m^4+3(\alpha')^2]\\
    +m^4[m^2(\varepsilon m^2r^2-1)(\alpha\ddot\alpha-3\dot\alpha^2)+\alpha^2(\alpha\ddot\alpha-\dot\alpha^2)]=0.
\end{multline}
The solution to this equation reduces to that given by Eq.\ \eqref{eq:SolforPhibyMM} for the static case if $\dot\alpha=0$.

The relevant equations of motion for $A$ are instead
\begin{subequations}\label{eq:EoMforAbyMM3}
    \begin{gather}
        (-m^2+\varepsilon m^4r^2+\alpha^2)\Phi'=0,\\
        (-m^2+\varepsilon m^4r^2+\alpha^2)\dot\Phi=0,\\
        \dot\alpha=0.\label{eq:EoMforAbyMM3-Birkhoff}
    \end{gather}
\end{subequations}
In particular, Eq.\ \eqref{eq:EoMforAbyMM3-Birkhoff} -- which corresponds to the equations of motion for the gauge potential $A$ with $(\mu,A,B)=(1,0,4)$ -- proves the pre-geometric Birkhoff's theorem in the MM theory for the class of solutions \eqref{eq:class-solutions2}. As before, from this result and the observations made right after Eqs.\ \eqref{eq:class-solutions2} and \eqref{eq:EoMforPhibyMM3}, it follows that the solutions to the full set of time-dependent equations of motion for $A$ in the MM theory (which includes but is not limited to Eqs.\ \eqref{eq:EoMforAbyMM3}) coincide with the solutions \eqref{eq:SolforAbyMM} presented for the static case, this time with $\Phi(t,r)$ undetermined in \eqref{eq:SolforAbyMMb} instead of $\Phi(r)$.

\section{Discussion and conclusions}\label{sec:conclusions}
The main result of this paper is finding pre-geometric configurations whose geometric analogue is the (anti-)de Sitter--Schwarzschild spacetime. These come in the form of exact solutions to the vacuum equations of motion of two different pre-geometric theories, namely the Wilczek and the MacDowell--Mansouri theories. After a mechanism of SSB near the Planck scale, a static and spherically symmetric black hole can thus emerge from a static and spherically symmetric pre-geometry of spacetime, with the (anti-)de Sitter--Schwarzschild metric effectively arising as an effect of the interaction of the pre-geometric gauge and Higgs-like fields in the spontaneously broken phase. The pre-geometric phase transition was shown to be inevitable in the Wilczek theory, as the gravitational Higgs field has to settle into its VEV under the hypothesis of static and spherical symmetry, which automatically leads to the SSB of the fundamental symmetry of spacetime and the emergence of gravity. In the case of the MacDowell--Mansouri theory, instead, the pre-geometric solution admits two branches, one being the same as the solution in the Wilczek theory and another one being the pre-geometric analogue of the (anti-)de Sitter spacetime with an undetermined Higgs-like field. This situation can be interpreted as a static and spherically symmetric pre-geometric configuration endowed with a pre-geometric (anti-)de Sitter core of Planckian size: as the gravitational Higgs field is free to evolve within the regular pre-geometric core, its SSB will eventually determine the emergence of the (anti-)de Sitter--Schwarzschild metric outside the core. The geometry of a static and spherically symmetric body is thus described by the (anti-)de Sitter--Schwarzschild spacetime right until a Planckian distance from the centre, wherein the unbroken phase of the fundamental pre-geometric theory is restored and the geometric singularity of Einstein gravity is removed. The expectation of getting rid of classical spacetime singularities thanks to quantum or emergent phenomena near the Planck scale is a hallmark of every quantum gravity approach, and finds a relatively simple realisation in the framework of Pre-geometric Gravity.

Due to the mathematical complexity of the pre-geometric field equations in vacuum, we did not address the uniqueness of the exact solutions found for pre-geometric black holes. This means that it is possible that other classes of solutions satisfying the same symmetry conditions may exist. The solutions computed in this article are those highlighted by the correspondence principle, i.e.\ those that are relevant for recovering known metric spacetimes in the spontaneously broken phase. It is conceivable, at least from a mathematical point of view, that additional pre-geometric configurations with spherical symmetry could be predicted by the same pre-geometric theories under consideration -- though this possibility would then beg the question about the physical character of the corresponding spacetime geometries in the metric phase.

As for the static nature of pre-geometric black holes with spherical symmetry, we were able to prove that it is a necessity for each exact solution that was examined. This result represents the pre-geometric Birkhoff's theorem, that is the pre-geometric analogue of Birkhoff's result for spherically symmetric black holes in GR. The only step away from completing this key consistency check between the pre-geometric and the metric theories rests, again, on a discussion of the uniqueness of the pre-geometric solutions (given that in GR \emph{any} metric with spherical symmetry is necessarily static in vacuum).

In-depth analyses of the phenomenological implications of pre-geometric black holes are left for future studies. Another important extension of the present work concerns the pre-geometry of more complicated black hole configurations, as well as exotic solutions such as wormholes. Along the lines of Refs.\ \cite{addazi:holographic,addazi:solution,addazi:generalized}, the pre-geometric formalism could also be adopted in order to delineate a novel pathway for the application of the holographic principle, which would certainly be of interest in the context of black hole physics and the quantum gravity program.

\acknowledgments
GM acknowledges the support of Istituto Nazionale di Fisica Nucleare (INFN), Sez.\ di Napoli, Iniziativa Specifica QGSKY. This publication is based upon work from COST Action CA21136 -- ``Addressing observational tensions in cosmology with systematics and fundamental physics (CosmoVerse)'', supported by COST (European Cooperation in Science and Technology).

\end{document}